# The role of transposable elements in the evolution of non-mammalian vertebrates and invertebrates

Noa Sela (noa.sela@bio.lmu.de)
Eddo Kim (kimedd@post.tau.ac.il)
Gil Ast (gilast@post.tau.ac.il)

# The role of transposable elements in the evolution of non-mammalian vertebrates and invertebrates

**Noa Sela[1,2], Eddo Kim[1] and Gil Ast[1]**

[1]Department of Human Molecular Genetics, Sackler Faculty of Medicine, Tel Aviv University, Tel Aviv 69978, Israel

[2] Currently at the Department Biology I, Ludwig-Maximilians-University Munich (LMU) Großhaderner Str. 2, D-82152, Planegg-Martinsried, Germany

Correspondence should be addressed to Gil Ast:

gilast@post.tau.ac.il.

**Emails:**

NS: noa.sela@bio.lmu.de

EK: kimedd@post.tau.ac.il

GA: gilast@post.tau.ac.il

**Abstract**

**Background:** Transposable elements (TEs) have played an important role in the diversification and enrichment of mammalian transcriptomes through various mechanisms such as exonization and intronization (the birth of new exons/introns from previously intronic/exonic sequences, respectively), and insertion into first and last exons. However, no extensive analysis has compared the effects of TEs on the transcriptomes of mammalian, non-mammalian vertebrates and invertebrates.

**Results:** We analyzed the influence of TEs on the transcriptomes of five species, three invertebrates and two non-mammalian vertebrates. Compared to previously analyzed mammals, there were lower levels of TE introduction into introns, significantly lower numbers of exonizations originating from TEs and a lower percentage of TE insertion within the first and last exons. Although the transcriptomes of vertebrates exhibit a significant level of exonizations of TEs, only anecdotal cases were found in invertebrates. In vertebrates, as in mammals, the exonized TEs are mostly alternatively spliced, indicating selective pressure maintains the original mRNA product generated from such genes.

**Conclusions:** Exonization of TEs is wide-spread in mammals, less so in non-mammalian vertebrates, and very low in invertebrates. We assume that the exonization process depends on the length of introns. Vertebrates, unlike invertebrates, are characterized by long introns and short internal exons. Our results suggest that there is a direct link between the length of introns and exonization of TEs and that this process became more prevalent following the appearance of mammals.

## Background

Transposable elements (TEs) are mobile genetic sequences that comprise a large fraction of mammalian genomes: 45%, 37% and 55% of the human, mouse and opossum genomes are made up of these elements, respectively [1-6]. TEs are distinguished by their mode of propagation. Short interspersed repeat elements (SINEs), long interspersed repeat elements (LINEs) and retrovirus-like elements with long-terminal repeats (LTRs) are propagated by reverse transcription of an RNA intermediate. In contrast, DNA transposons move through a direct 'cut-and-paste' mechanism [7]. TEs are not just "junk" DNA but rather are important players in mammalian evolution and speciation through mechanisms such as exonization and intronization [8-11]. Alternative splicing of exonized TEs can be tissue specific [12, 13] and exonization contributes to the diversification of genes after duplication [14].

Most exonized TEs are alternatively spliced, which allows the enhancement of transciptomic and proteomic diversity while maintaining the original mRNA product [9-11, 15, 16]. Exonization can take place following insertion of a TE into an intron. However, the majority of invertebrate introns are relatively short [17] and are under selection to remain as such due to the intron definition mechanism by which they are recognized [18-21]. Thus, there is presumably a selection against TE insertion into such introns. However, with the presumed transition from intron to exon definition during evolution [20, 22], introns were freed from length constraints. This reduced the selection against insertion of TE into introns and a large fraction of mammalian introns contain TEs, although only a small fraction are exonized [16]. For the most part, TEs have not been inserted within internal coding exons; they are found in first and last exons and in untranslated regions (UTRs), apparently the outcome of coding constraints [16].

The impact of TEs on the genomes of human [8-11, 16, 23-26], dog [4, 5], cow [3], mouse [16] and opossum [6, 27] has been extensively studied. Bejerano and colleagues have shown that SINE elements that were active in non-mammalian vertebrates during the Silurian period are the source of ultra-conserved elements within mammalian genomes [28]. However, with this exception there have been no systematic large-scale analyses of the impact of TEs on the transcriptomes of non-mammalian genomes. To address this issue we compiled a dataset of all TE families in the genomes of chicken (*Gallus gallus*), zebrafish (*Danio rerio*), sea squirt (*Ciona intestinalis*), fruit fly (*Drosophila melanogaster*) and nematode (*Caenorhabditis elegans*). We examined the location of each TE with respect to annotated genes. We found that the percentage of TEs within transcribed regions of these non-mammalian vertebrates and invertebrates is much lower than the percentage observed within mammals. We also found evidence for TE exonization in all species we examined. However, the magnitude of this process differed among the tested organisms; we detected a substantially higher level of exonizations in vertebrates (*G. gallus* and *D. rerio*) compared to invertebrates (*D. melanogaster* and *C. elegans*). There is a higher abundant of TEs in intronic sequences and introns are much larger in vertebrates than in invertebrates, suggesting that TEs located in long introns provide fertile ground for testing new exons via the exonization process. Overall, the results we present suggest that TE exonization is a mechanism for transcriptome enrichment not only in mammals, but also in non-mammalian vertebrates as well as in invertebrates, albeit to a lesser extent.

## Results

### Genome-wide analysis of TE insertions within the transcriptomes of five non-mammalian species

To evaluate the effect of TEs on the transcriptomes of non-mammals, we analyzed the genomes of five non-mammalian vertebrates and invertebrates: *Gallus gallus*, *Danio rerio*, *Ciona intestinalis, Drosophila melanogaster* and *Caenorhabditis elegans*. To calculate the total number of TEs in each genome, the number of TEs in introns, and the number of TEs present within mRNA molecules, we downloaded EST and cDNA alignments and repetitive element annotations for these five genomes from the University of California Santa Cruz (UCSC) genome browser [24] (see Materials and Methods and also [29]). Tables 1, 2, 3, 4 and 5 summarize our analyses for each of these species.

TEs have altered the transcriptomes of mammals and the examined non-mammalian genomes differently. First, the portion of the genome covered by TEs differs dramatically. In mammalian genomes, TEs occupy between 37% and 52% of the genome [1-6, 30]. In the five evaluated non-mammalian genomes, TEs account for approximately 10% of the genome sequence, with the exception of *Danio rerio*, where TEs occupy 26.5% (Figure 1). The second important difference is related to the types of TEs observed. In mouse and human, SINEs are the most abundant TEs. In the *Gallus gallus* genome, LINEs (belonging to the family of CR1 repeats) account for 79% of all TEs. In the *Danio rerio* genome, more than 75% of TEs are DNA transposons; whereas in *D. melanogaster*, LTRs are the most abundant TEs accounting for 44% of the elements observed. Finally, DNA transposons account for 95% of TEs in *C. elegans*. These differences have influenced the transcriptomes of non-mammals: in contrast to SINEs, which are non-autonomous mobile elements that do not encode for proteins, all other families of TEs are autonomous and contain at least one open reading frame.

**Insertion of TEs within intronic sequences**

Deeper analysis of the non-mammalian genomes revealed that TEs are less likely to be fixed within transcribed regions relative to orthologous regions in human and mouse [16]. In *Gallus gallus*, *Danio rerio* and *Ciona intestinalis*, 33.2%, 47.3% and 39.4% of TEs reside within introns, respectively, whereas in the human genome, ~60% of TEs reside within introns [16] ($\chi^2$, p-value = 0, for a comparison of TEs either in *Gallus gallus*, *Danio rerio*, or *Ciona intestinalis*, versus human). In the genome of *Drosophila melanogaster*, the fraction of intronic TEs is 60%, similar to that of mammals ($\chi^2$, p-value = 0.3 compared with human); in *C. elegans* 53% of TEs reside within intronic sequences, significantly lower compared to human ($\chi^2$, p-value = 1.1e-42). Among all TEs, LTRs have the lowest insertion levels within intronic sequences compared to other TE families in all genomes analyzed (Tables 1, 2, 3, 4, and 5) as was also observed for human and mouse [16]. The lower level of invasion of TEs within intronic sequences in *D. melanogaster* may be due in part to the fact that a large fraction of TEs in *Drosphila* are LTR sequences that have a lower tendency than other TE families to reside within introns [16, 31].

We next evaluated the TE distribution and determined the length of introns that contain TEs (Figure 2). We analyzed all intronic sequences of human (total of 184,145 introns), mouse (total of 177,766 introns), *Gallus gallus* (total of 167,626 introns), *Danio rerio* (total of 194,221 introns), *Ciona intestinalis* (total of 34,328 introns), *Drosophila melanogaster* (total of 41,145 introns) and *C. elegans* (total of 98,695 introns) for TE insertions to determine the percentage of TE-containing introns (Figure 2A). The fraction of the introns that contain TEs in the non-mammalian vertebrates *Gallus gallus* and *Danio rerio* is 21.3% and 44.3%, respectively, substantially lower than that of mammals (63.4% and 60.2% in human and mouse, respectively). The fraction of introns containing TEs in the deuterostome *C. intestinalis* is 33.4%, very similar to the percentage in non-mammalian vertebrates. In contrast, the fraction of introns that contain TEs in invertebrates *Drosophila melanogaster* and *C. elegans* is 1.7% and 5.6%, respectively. These results indicate that only a very small portion of introns in invertebrates contain TEs (2-5%) compared to 20-40% of introns in non-mammalian vertebrates and ~60% in mammals.

We also examined the average length of introns containing TEs. In *C. elegans* the median length of an intron containing a TE is ~700 bp (after subtracting TE length, the median intron size is 477 bp), compared to ~3000 bp in human, mouse, chicken and zebrafish. The median length of introns that contain TEs in the fruit fly is around 6000 bp (after subtracting the TE length, the median intron length is 5822 bp), whereas the median length of introns in fruit fly is only 72 bp [17] (Figure 2B and 2C). Therefore, the introns in fruit fly that contain TEs are presumably under different selective pressure than the vast majority of introns in this organism; we assume that these TE-containing introns are not selected via the intron definition mechanism [19]. In general, we found a positive correlation between the fraction of introns containing

TEs and median length of introns (Figure 2C), implying that TE insertions have played a role in the evolution of intron size.

Previous analysis of human and mouse transcriptomes revealed that there is a biased insertion and fixation of some families of TEs within intronic sequences [16]: L1 and LTRs are most often fixed in their antisense orientation relative to the mRNA molecule. Our current analysis also revealed a bias toward antisense fixations of LTR sequences within *G. gallus*, *D. rerio* and *D. melanogaster* genomes (Supplementary table S1 in Additional file 1). This biased insertion is also correlated with a lower tendency of LTRs to reside within intronic sequences relative to other families of TEs (see Tables 1, 2, 3, 4 and 5 for data on non-mammalian genomes and [16] for data on human and mouse). A bias toward antisense orientation was also observed for DNA transposons in *G. gallus* and *D. melanogaster* and for LINEs in *D. melanogaster*. These biased insertions are presumably due to potential for co-transcription of TEs that already contain coding sequences. Insertion in a sense orientation would introduce another promoter into the transcribed region, which is likely to be deleterious and therefore selected against.

**Exonizations within vertebrates and invertebrates**

In mammals, new exonizations resulting from TEs are mostly alternatively spliced cassette exons [10, 11, 15, 16, 26, 32, 33]. In non-mammalian genomes, the level of alternative splicing is lower than that of mammals, with the exception of chicken where levels of alternative splicing are comparable to those in human [34]. We analyzed the splicing patterns of the TE-derived exons in the four non-mammalian species that contain TE-derived exons; the analysis was based on alignment data between EST/cDNA sequences and their corresponding genomic regions. The TE-derived exons in *Danio rerio*, *C. intestinalis* and *C. elegans* were predominantly alternatively spliced (Figure 3), a phenomenon similar to that found in mammals, suggesting that similar evolutionary constraints (reviewed in [22, 26, 35]) affect exonizations of mammals and species outside the mammalian class. In *D. melanogaster*, there are no exonized TEs in which one of the splice site results from the TE sequence. *G. gallus* is an exception: In this species many TE exonizations were constitutively spliced. However, this observation may be a result of a substantially lower number of ESTs available for *G. gallus* (Supplementary table S2 in Additional file 2). Without sufficient EST data, identification of alternatively spliced exons is difficult and exons may be mistakenly classified as constitutively spliced. We will need to re-evaluate this statement once additional EST coverage becomes available for *G. gallus*.

The majority of TE exonizations occur in genomic loci that are not annotated as genes by the RefSeq [36, 37] or Ensembl [38, 39] databases. It may be that these genes are species-specific and are not annotated due to a lack of homologs; alternatively, these may be non-protein coding genes. Of the exonizations found in annotated genes, 66-87% are found within the coding sequence (Supplementary table S3 in Additional file 3). Exonizations in non-mammals frequently disrupted the open reading frame of a protein, similar to results previously reported for human and mouse. In *G. gallus*, *D. rerio* and *C. intestinalis* only 38% to 50% of the exonized TEs have lengths divisible by three and therefore maintain the original coding sequence (Supplementary table S3 in Additional file 3).

In *D. melanogaster*, we found no evidence for exonizations using current ESTs or cDNA. We did identify three cases in which TEs were inserted into internal exons, all within the coding sequence (see Figure 4, and also Supplementary text S1 in Additional file 4 for exon sequences). In these cases, the length of the inserted TEs (LINEs) was found to be divisible by three and the sequences did not contain stop codons. Thus, the insertion of these TEs into the coding exons did not alter the reading frame of the downstream exons, but rather added new amino acid sequence to the proteins. These insertions result in extremely long exons (668, 2025 and 4077 bp). One of these exons is flanked by very short introns (82 and 68 bp for the upstream and downstream introns, Figure 4C) and two are flanked by a short downstream intron and a long upstream intron (85 and 70 bp for the downstream introns and 1003 and 689 bp for the upstream introns, Figure 4 A and B). In mammals, no evidence was found for TE insertions into coding exons [15, 16]. We assume that this difference between mammals and *Drosophila* is due to the fact that in *D. melanogaster* the intron definition mechanism is dominant, which allows the lengthening of exons in a short-intron environment [19].

We have recently shown evidence for transduplication of protein coding genes within DNA transposons in *C. elegans* [40]. In this analysis, we found that DNA transposons have also influenced the coding sequence of *C. elegans* genes by means of exonization. One such example is an alternatively spliced exon of 73 bp in the CDS of a hypothetical protein (Y71G12A.2). The accession number of the RefSeq sequence that contains the exonization is [NM_058514]; the accession number of the RefSeq sequence without the exonization is [NM_001129082] (both RefSeq mRNA sequences have been reviewed). The gene is conserved within nematodes (*C. remanei,*

*C. briggsae, C. brenneri and C. japonica*). It should be noted that only a single *C. elegans* individual has been sequenced and this event might be restricted to this individual. However, this event does suggest that an exonization mechanism operates in nematodes.

New exonizations resulting from TEs were found in non-vertebrate deuterostome *C. intestinalis* (9 exonizations, Table 3) and in much larger quantities in vertebrates (70 in *G. gallus* and 253 in *D. rerio*, Tables 1 and 2, respectively). The number of exonizations was not directly correlated to the number of ESTs available for each genome, suggesting that our results reflect a true difference in the extent of exonization across organisms. There are 599,785 ESTs for *G. gallus*, 1,380,071 ESTs for *D. rerio*, 1,205,674 ESTs for *C. intestinalis*, 573,981 ESTs for *D. melanogaster* and 352,044 ESTs for *C. elegans* (Table S4 in Additional file 5). The majority of exonizations found in *G. gallus* result from the CR1 LINE element, which is the most abundant TE within the *G. gallus* genome.

In the zebrafish genome, like that of mammals, the most abundant TEs are SINE elements. About 68% (77,436 copies) of zebrafish TEs are intronic SINEs that belong to the HE1 family of SINEs; these HE1 SINEs comprise almost 10% of the zebrafish genome [41]. The HE1 are tRNA-derived SINEs with a consensus sequence of 402 bp long found also in elasmobranches (the subclass of cartilaginous fish) [42]. The HE1 family is the oldest known family of SINEs, dated to 200 million years ago [42]. The HE1 SINEs were previously shown to be the source of mutational activity in the zebrafish genome and have been used as a tool for characterization of zebrafish populations [41]. SINEs have resulted in a substantial number of new exons (135 exons, Table 2) and that 84.4% (114 exons) are derived from HE1 SINEs. Of the 114 cases of exonizations from HE1 elements, 69 insertions were in the sense orientation and 45 in the antisense orientation with respect to the coding sequence. These results suggest that there is no statistical preference for exonization in a specific orientation ($\chi^2$, p-value = 0.14). A typical SINE contain a poly(A) tail. Most of exonizations originated from SINEs (*Alu,* B1, MIR) are from elements inserted into introns in the antisense orientation, relative to the coding sequence [10, 15, 16]. When SINE with poly(A) insert into introns in the antisense orientation the poly(A) tail becomes a poly(U) in the mRNA precursor and thus can serve as a polypyrimidine tract for mRNA splicing [9]. The no preference for exonization in a specific orientation of HE1 in zebrafish is presumably because of the absence of poly(A) tail from the sequence of this SINE [43]. The tRNA-related, 5'-conserved regions of the HE1 element contain sequences that serve as 3' and 5' splice sites (Figure 5A). When a sense HE1 region is exonized, the exonization is within the 5' conserved area, whereas exonizations from HE1 elements in the antisense orientation encompass the entire HE1 sequence (Figure 5). Finally, DNA repeat elements are also substantial contributors of new exons in zebrafish (109 exons, Table 2). The exonization of DNA repeats is not biased to one of the orientations ($\chi^2$, p-value = 0.13).

**TE insertions into the first and last exons**

Our analysis shows that the influence of TEs on the transcriptomes of non-mammals is not limited to the creation of new internal exons: TEs also modified the mRNA by insertion into the first or last exon of a gene. This type of insertion causes an elongation of the first or last exons and usually affects the UTR (Figure 4B). In human, this type of insertion has been shown to create new non-conserved polyadenylation signals [44], influence the level of gene expression [45] and create new microRNA targets [46, 47].

For the analysis of the number of TE insertions within the first or last exons in chicken, zebrafish, fruit fly and nematode, we used the UCSC annotated RefSeq genes and examined those full-length sequences in which the entire transcript is annotated and a consensus mRNA sequence exists (refGene table). Our results indicate that TEs occupy a lower percentage of the base pairs within the first and last exons in mouse, chicken, zebrafish, *C. intestinalis*, *D. melanogaster* and *C. elegans* than do TEs in human first and last exons (see Table S4 in Additional file 5 and Table S5 in Additional file 6). Our previous analysis showed that in human annotated genes, the average lengths of the first and last exons are 465 and 1,300 bp, respectively, and in mouse genes the first exon has an average length of 393 bp and the last exon an average length of 1,189 bp [16]. The average lengths of the first and last exons in the non-mammalian species are shown in Figure 6 (see also Supplementary table S4 in Additional file 5 and Table S5 in Additional file 6); all have average exon lengths shorter than those of human and mouse. The fly has on average the longest first exons among the non-mammalian species, whereas the chicken genome contains the longest last exons on average (Figure 6).

## Discussion

In this study, we examined the influence of TEs on the transcriptomes of five species, including two vertebrates, one non-vertebrate deuterostome and two invertebrates. We compared our data to previous results generated for two mammalian species (human and mouse) [16]. We observed significant differences between vertebrates and invertebrates regarding the exonizations that have resulted from TE insertion. In chicken and zebrafish, we found dozens of exonizations: 70 exons were a result of TE insertions in *G. gallus* and 153 in *D. rerio*. Lower on the evolutionary tree, TEs were much less frequently exonized, if at all. In the deuterostome *C. intestinalis*, we found only 12 exons that resulted from TEs and none were observed in *D. melanogaster* and *C. elegans*.

The prevalence of exonizations within human and mouse (around 1800 new exons in human and around 500 new exons in mouse [16]) is mainly attributed to the existence of very large introns and the dominance of the exon definition mechanism for splice site selection in mammals [48]. Invertebrates, in contrast, have short introns and long exons [17]. The transition from the intron definition mechanism used by invertebrates to that of exon definition during evolution presumably reduced selective pressure on intron length, which probably allowed insertion of TEs into intron sequences without deleterious consequences [48, 49]. As could be expected due to the difference in the length of introns, the numbers of TEs located in intron sequences is substantially lower in the non-mammalian genomes compared to mammalian genomes. One might expect that in organisms where the splicing machinery functions via the intron definition mechanism, insertion of TEs into the longer coding exons would be prevalent. However, only three cases of such insertions were detected in the *D. melanogaster* genome, suggesting that this mechanism of transcriptome enrichment is evolutionary unfavorable. It is likely that TE insertions into coding exons are not propagated as these events would alter the coding sequence immediately upon insertion. A previous genome-wide analysis of TEs in *Drosophila* and their association with gene location found a small number of fixed TEs [50]. However other analyses have shown that TEs have played an important role in adaptation of fruit flies [51]. One of the most significant reports was that of the truncation of the CHKov1 gene by a TE leading to resistance to pesticides [52].

SINEs and LINEs were shown in many publications to be good substrates for the exonization process because of their special structure [9, 11, 15, 16, 26]. In mammalians and other vertebrates higher level of SINEs and LINEs within intron sequences gave rise to a greater level of exonization due to the pre-existence of splice site like sequences, such as the polypyrimidine tract and putative 5' splice sites [9, 11, 15, 16, 26].

TEs are often inserted into exonic regions that are part of UTRs. Our analysis indicated that, on average, the size of the last exons is longer in mammals compared to vertebrates and more so in invertebrates. The differences in the length of the last exons are correlated with an increase in the percentage of TEs inserted into last exons. Insertions of TEs into UTRs may alter levels of gene expression, create new targets for microRNA binding, or even result in precursors for new microRNAs [46, 47, 53]. The presumably increase in the size of last exons and in the percentage of TEs within these exons from invertebrates to mammals may have led to the high level of regulatory complexity observed in high organisms. Exonization of TEs is wide-spread in mammals, less so in non-mammalian vertebrates, and very low in invertebrates.

## Conclusions

Our results suggest that there is a direct link between the length of introns and exonization of TEs and that this process became more prevalent following the appearance of mammals.

## Materials and methods

### Dataset of TEs within coding regions of five species

Chicken (galGal3, May 2006), zebrafish (danRer4, March 2006), fruit fly (dm2, April 2004), *C. elegans* (ce2, March 2004) and sea squirt (ci2, March 2005) genome assemblies were downloaded, along with their annotations, from the UCSC genome browser database [24, 54]. EST and cDNA mappings were obtained from chrN_intronEST and chrN_mrna tables, respectively. TE mappings data were obtained from chrN_rmsk tables and TE sequences were retrieved from genomic sequences using the mapping data. A TE was considered intragenic if there was no overlap with ESTs or cDNA alignments; it was considered intronic if it was found within an alignment of an EST or cDNA defined as an intronic region. Finally, a TE was considered exonized if it was found within an exonic part of an EST or cDNA (except the first or last exon of the EST/cDNA), and possessed canonical splice sites. Next, we associated the intronic and exonized TEs with genomic positions of protein-coding genes by comparisons with RefSeq [55] gene tables from the UCSC table browser [54]. Positions of the TE hosting intron/exon and the mature mRNA were calculated using the gene tables. Association of the gene to the mRNA and protein accessions and to descriptions from RefSeq and Swiss-Prot was done through the kgXref and refLink tables in the UCSC genome browser database [54]. All data used has been published [22, 29].

**Analysis of retroelement insertions within the first and last exons and assessment of untranslated region fraction in known genes**

The tables refGene and refLink were used to examine the relative lengths of the UTRs and the coding sequences (CDSs) within chicken, zebrafish, sea squirt, fruit fly and nematode genes and to find the first and last exons. The analysis of TE content was done using the RepeatMasker software [38] and repbase [56, 57].

**Estimation of the fraction of TEs within introns**

We determined the TE fraction within intronic sequences using the UCSC genome browser and GALAXY [54, 58, 59]. Introns of chicken (*G. gallus*, Build 1.1), zebrafish (*D. rerio*, release Zv4), *C. elegans* (Release 2003) and *D. melanogaster* (Build 4.1) were extracted from the Exon-Intron Database [60], [61]. When alternatively spliced isoforms of the same gene were present, only the first annotated isoform was extracted; all other isoforms were excluded in order to avoid redundancy. The analysis of the TE content was done using RepeatMasker software and repbase [56, 57]. In the case of *C. intestinalis*, the analysis of 34,328 intronic sequences was done using the GALAXY server [59] and UCSC genome browser tables [54].

**Statistical analysis**

For the comparative analysis of insertions within introns of various species we used a contingency table $\chi^2$ test. In cases where the contingency table was a 2×2 table, the Fisher's exact test was used. To assess the tendency of exonizations to occur within UTRs we used the goodness-of-fit $\chi^2$ test. The null hypothesis was the fraction of the UTR and CDS within the RefSeq gene list of chicken, zebrafish, sea squirt, fruit fly and *C. elegans*. The calculation of *p* values for differences between two populations was measured according to the data distribution. The Kolmogorov-Smirnov test was used to test for normal distribution. The t-test was used to calculate statistical differences.

**Abbreviations**

TE: transposable element; LTR: long interspersed repeat; SINE: short interspersed element; LINE: long interspersed element; MIR: mammalian interspersed repeat; CDS: coding sequence; UTR: untranslated region; EST: Expressed sequence tag; cDNA: complementary DNA; Refseq: reference sequence.

**Authors' contributions**

NS carried out the computational analysis. NS and GA conceived of the study. EK gave professional advice regarding interpretation of results. NS, EK and GA drafted the manuscript.

**Acknowledgments**

The authors thank Wojciech Makalowski and Gyorgy Abrusan for stimulating discussions. This work was supported by the Cooperation Program in Cancer Research of the Deutsches Krebsforschungszentrum (DKFZ) and Israel's Ministry of Science and Technology (MOST) and by a grant from the Israel Science Foundation (40/05), ICRF, DIP and EURASNET. NS is supported by the LMU excellence fellowship.

## Figure legends

**Figure 1: Non-mammalian vertebrate and invertebrate genomes have lower levels of TEs than mammalian genomes.** Evolutionary trees for chicken [30], zebrafish, sea squirt [62], *Drosophila* [63] and worm [63]. Percentages of TEs in each genome are shown on the right.

**Figure 2: The fraction of introns containing TEs and their median lengths in non-mammalian and mammalian transcriptomes.** (**A**) The fraction of TE-containing introns within five non-mammalian genomes compared to that of human and mouse (for details see Materials and Methods). (**B**) A graph of the median length of introns containing TEs compared to that of introns without TEs (marked in grey and black, respectively) in the different organisms. (**C**) Positive correlation between median intron length and the fraction of TEs containing introns. Intron lengths were taken from [17].

**Figure 3: The effect of TEs on non-mammalian transcriptomes.** (**A**) Summary of the number of exonized TEs in the different species. (i) Illustration of the exonization process, in which a TE (gray box) was inserted into an intron (line). Exonization of TE may (ii) generate a cassette exon, (iii) create an alternative 5' splice site, (iv) create an alternative 3' splice site, or (v) be constitutively spliced. The table on the right shows the numbers of exonized TEs in each of the examined species. (**B**) Summary of the effect of TE insertions into the first or last exons. (i) Illustration of insertion of TEs (gray box) into an exon (white box). The insertion of the TEs may enlarge (ii) the first or (iii) the last exon.

**Figure 4: Three cases of TE insertions into internal exons in *Drosophila melanogaster*.** Schematic representations of TE insertions into *Drosophila* internal exons. White boxes and lines represent exons and introns, respectively. The grey boxes show insertion of TEs into exons. The TE family is indicated beneath the gray box, along with the length of each inserted TE. Lengths of the introns and exons flanking the inserted exon are indicated. Genes with insertions are (**A**) cno, (**B**) CG14821 and (**C**) nej.

**Figure 5: The HE-1 SINE exonization in zebrafish.** (**A**) Alignment of HE1 SINE from *Danio rerio* and HE1 SINE from bullhead shark showing the different sections within the transposable element according to [43]. The letters y and r denote pyrimidine and purine, respectively. (**B**) Non-redundant distribution and orientation of exonized HE1 SINE sequences in which both the 5' and the 3' splice sites are within the HE1 SINE sequence. The exonized HE1 SINE sequence regions are aligned against an HE1 SINE consensus element. Each line is a different EST showing exonizations and the box in the middle represents the HE1 element. The number of cases that select that site as a 5' splice site or as a 3’ splice site are in red and blue, respectively. Exonizations in the sense and antisense orientations are shown above and below the schematic representation of the HE1.

**Figure 6: Average lengths of first and last exons compared to the fraction of TEs inserted into exons.** (**A**) The Y axis indicates average length of first exon in the six examined organisms (bars) and the percentage of base pairs that originated from TEs (line). (**B**) Similar analysis for last exons. Note that the Y axes are different in scale.

## Tables

**Table 1: Transposable elements in *Gallus gallus***

| TE | Total | Intronic | TEs in introns within RefSeq | TEs in introns of non-RefSeq | Exons within RefSeq alignments* | Exons in non-RefSeq alignments** |
|---|---|---|---|---|---|---|
| SINE | 27 | 10 (37%) | 1 | 9 | 0 | 0 |
| LINE | 188,302 | 65,035 (34.5%) | 14,482 | 50,553 | 8 | 45 |
| LTR | 28,719 | 7553 (26.3%) | 1501 | 6052 | 0 | 8 |
| DNA | 20,808 | 6554 (31.4%) | 1446 | 5108 | 1 | 8 |
| **Total** | **237,856** | **79,152 (33.2%)** | **17,430** | **61,722** | **9** | **61** |

* Number of exons found within annotated RefSeq genes.
** Number of exons for which ESTs are not found within annotated RefSeq genes.

**Table 2: Transposable elements in *Danio rerio***

| TE | Total | Intronic | TEs in introns within RefSeq | TEs in introns of non-RefSeq | Exons within RefSeq alignments* | Exons in non-RefSeq alignments** |
|---|---|---|---|---|---|---|
| SINE | 259,684 | 113,926 (43.9%) | 46,679 | 67,247 | 14 | 121 |
| LINE | 80,412 | 37,228 (46.3%) | 14,671 | 22,557 | 2 | 4 |
| LTR | 53,028 | 21,496 (40.5%) | 6761 | 14,735 | 2 | 1 |
| DNA | 1,208,155 | 585,408 (48.4%) | 257,438 | 327,970 | 37 | 72 |
| **Total** | **1,601,279** | **758,058 (47.3%)** | **325,549** | **432,509** | **55** | **198** |

* Number of exons which their ESTs are found within annotated RefSeq genes.
** Number of exons which their ESTs are not found within annotated RefSeq genes.

**Table 3: Transposable elements in *Ciona intestinalis***

| TE | Total | Intronic | TEs in introns within RefSeq | TEs in introns of non-RefSeq | Exons within RefSeq alignments* | Exons in non-RefSeq alignments** |
|---|---|---|---|---|---|---|
| SINE | 51,021 | 20,360 (39.9%) | 826 | 19,534 | 0 | 3 |
| LINE | 29,369 | 11,172 (38%) | 493 | 10,679 | 0 | 0 |
| LTR | 491 | 112 (22.8%) | 2 | 110 | 0 | 0 |
| DNA | 55,300 | 22,056 (39.9%) | 1025 | 21,031 | 0 | 9 |
| **Total** | **136,181** | **53,700 (39.4%)** | **1851** | **51,849** | **0** | **12** |

* Number of exons which their ESTs are found within annotated RefSeq genes.
** Number of exons which their ESTs are not found within annotated RefSeq genes.

**Table 4: Transposable elements in *Drosophila melanogaster***

| TE | Total | Intronic | TEs in introns within RefSeq | TEs in introns of non-RefSeq | Exons within RefSeq alignments* | Exons in non-RefSeq alignments ** |
|---|---|---|---|---|---|---|
| SINE | 0 | 0 | 0 | 0 | 0 | 0 |
| LINE | 4755 | 2964 (62%) | 1258 | 1706 | 0 | 0 |
| LTR | 10,259 | 5394 (52%) | 2014 | 3380 | 0 | 0 |
| DNA | 8028 | 5560 (69%) | 3231 | 2329 | 0 | 0 |
| **Total** | **23,042** | **13,918 (60%)** | **6503** | **7415** | **0** | **0** |

* Number of exons which their ESTs are found within annotated RefSeq genes.
** Number of exons which their ESTs are not found within annotated RefSeq genes.

**Table 5: Transposable elements in *Caenorhabditis elegans***

| TE | Total | Intronic | TEs in introns within RefSeq | TEs in introns of non-RefSeq | Exons within RefSeq alignments* | Exons in non-RefSeq alignments** |
|---|---|---|---|---|---|---|
| SINE | 524 | 243 (46%) | 230 | 13 | 0 | 0 |
| LINE | 428 | 103 (24%) | 90 | 13 | 0 | 0 |
| LTR | 606 | 137 (22%) | 126 | 11 | 0 | 0 |
| DNA | 32,977 | 17,724 (53%) | 17,175 | 549 | 4 | 0 |
| **Total** | **34,535** | **18,207 (53%)** | **17,621** | **586** | **4** | **0** |

* Number of exons which their ESTs are found within annotated RefSeq genes.
** Number of exons which their ESTs are not found within annotated RefSeq genes.

## Additional files

**Additional file 1: Table S1.** Number of sense/antisense TE insertions within intronic sequences.

**Additional file 2: Table S2.** Number of exonizations versus number of ESTs.

**Additional file 3: Table S3.** Number of exonizations found in the CDS or UTR (within annotated genes).

**Additional file 4: Text S1.** Sequences of the internal *Drosophila melanogaster* exons with TE insertions.

**Additional file 5: Table S4.** Average lengths and average TE percentages in first exons.

**Additional file 6: Table S5.** Average lengths and average TE percentages in the last exons.

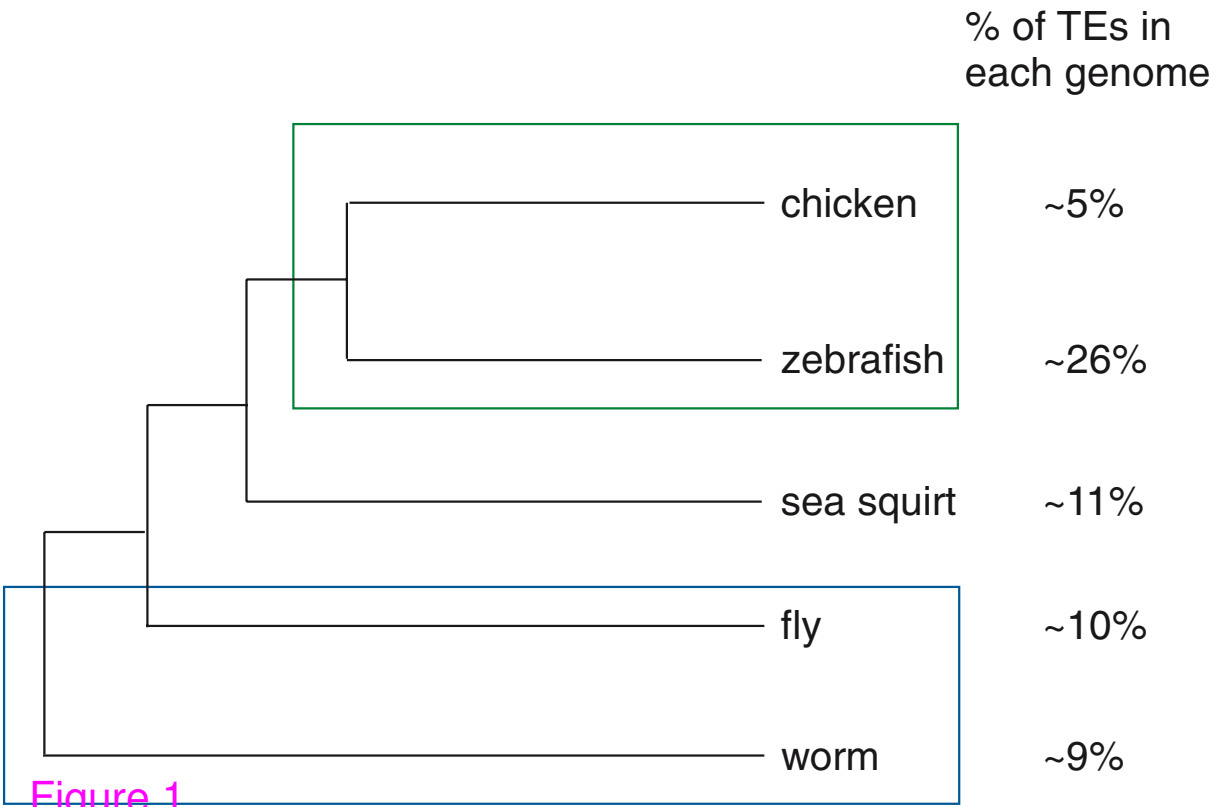
Figure 1

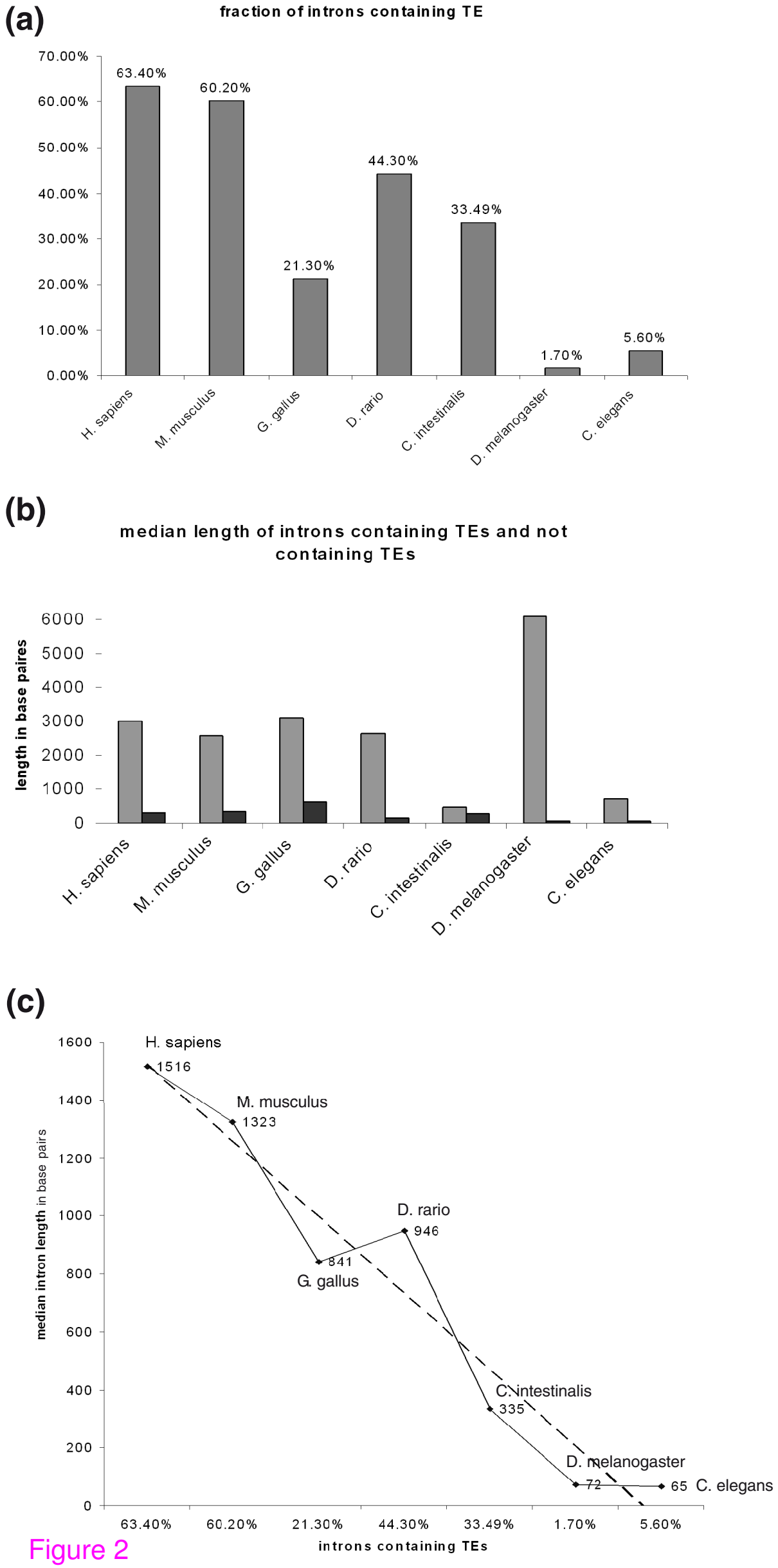

(a)
fraction of introns containing TE
70.00%
60.00%
50.00%
40.00%
30.00%
20.00%
10.00%
0.00%
63.40%
60.20%
21.30%
44.30%
33.49%
1.70%
5.60%
H. sapiens
M. musculus
G. gallus
D. rario
C. intestinalis
D. melanogaster
C. elegans
(b)
median length of introns containing TEs and not containing TEs
length in base paires
6000
5000
4000
3000
2000
1000
0
H. sapiens
M. musculus
G. gallus
D. rario
C. intestinalis
D. melanogaster
C. elegans
(c)
median intron length in base pairs
1600
1400
1200
1000
800
600
400
200
0
H. sapiens
1516
M. musculus
1323
D. rario
946
841
G. gallus
C. intestinalis
335
D. melanogaster
72
65
C. elegans
63.40%
60.20%
21.30%
44.30%
33.49%
1.70%
5.60%
introns containing TEs

Figure 2

**(a)**

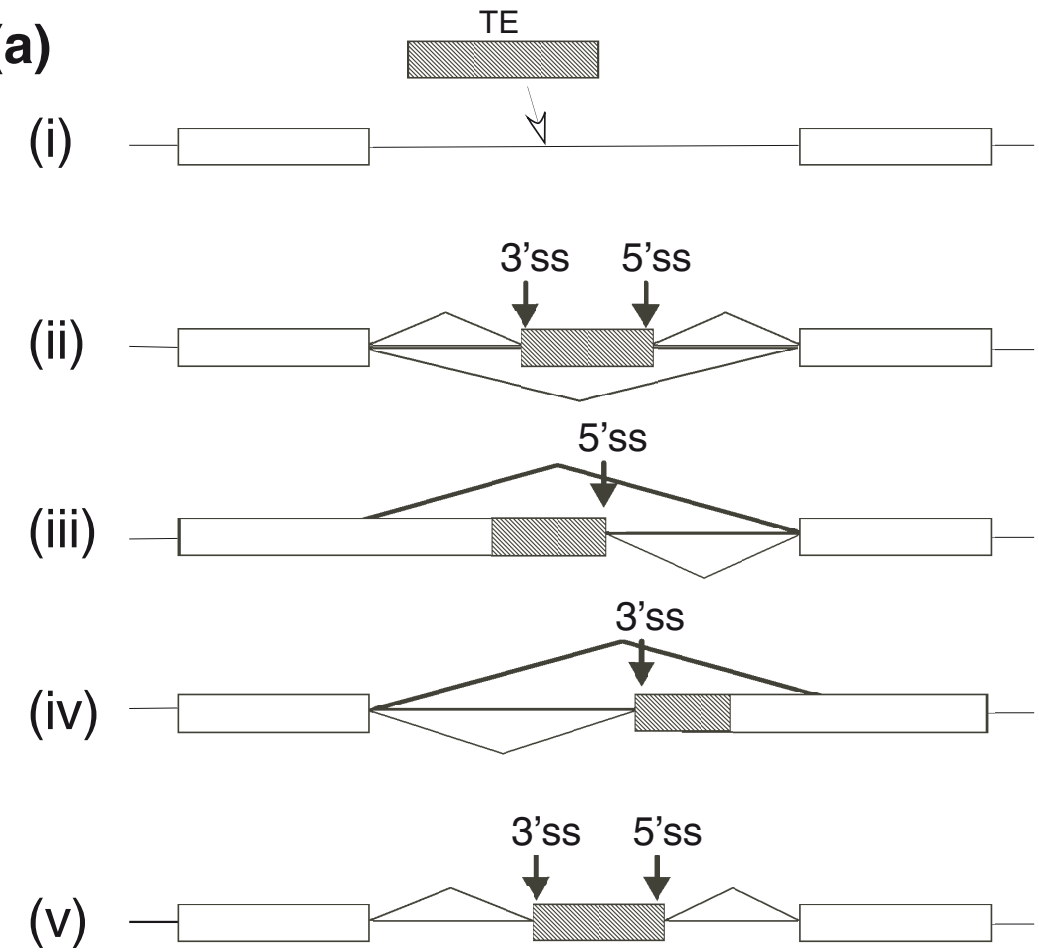

| | vertebrates | | | invertebrates | |
|---|---|---|---|---|---|
| species | Chicken | Zebrafish | C. intestinalis | D. melanogaster | C.elegans |
| Alt. Skip | 16 | 149 | 9 | 0 | 1 |
| Alt. 5'ss | 0 | 0 | 1 | 0 | 1 |
| Alt. 3'ss | 0 | 3 | 1 | 0 | 1 |
| Const. | 26 | 100 | 1 | 0 | 1 |

**(b)**

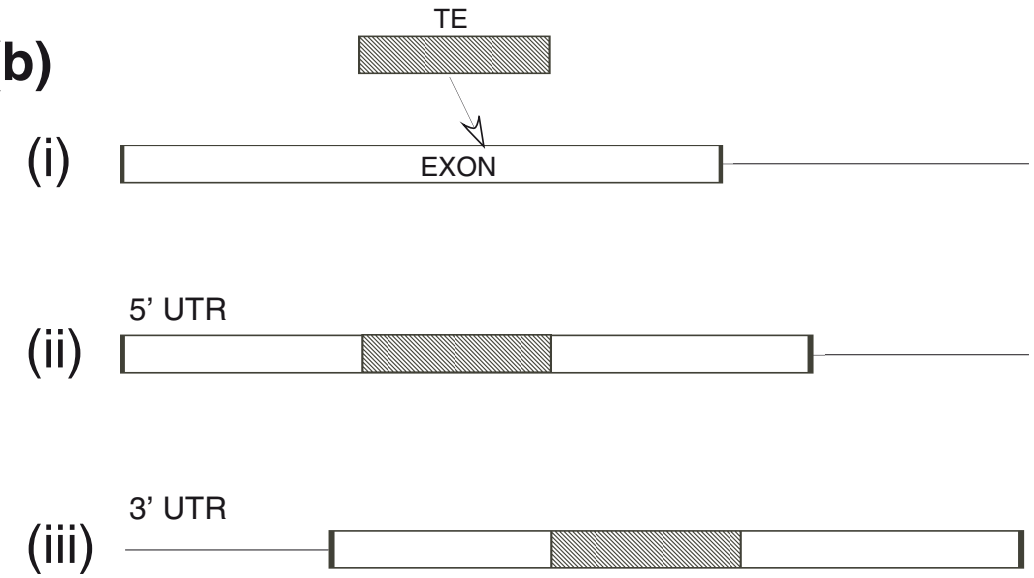

| | vertebrates | | | invertebrates | |
|---|---|---|---|---|---|
| species | Chicken | Zebrafish | C. intestinalis | D. melanogaster | C.elegans |
| insertion 5UTR | 0.11% | 0.34% | 0.24% | 0.29% | 0.6% |
| insertion 3UTR | 0.39% | 3.37% | 1.32% | 0.37% | 0.79% |

Figure 3

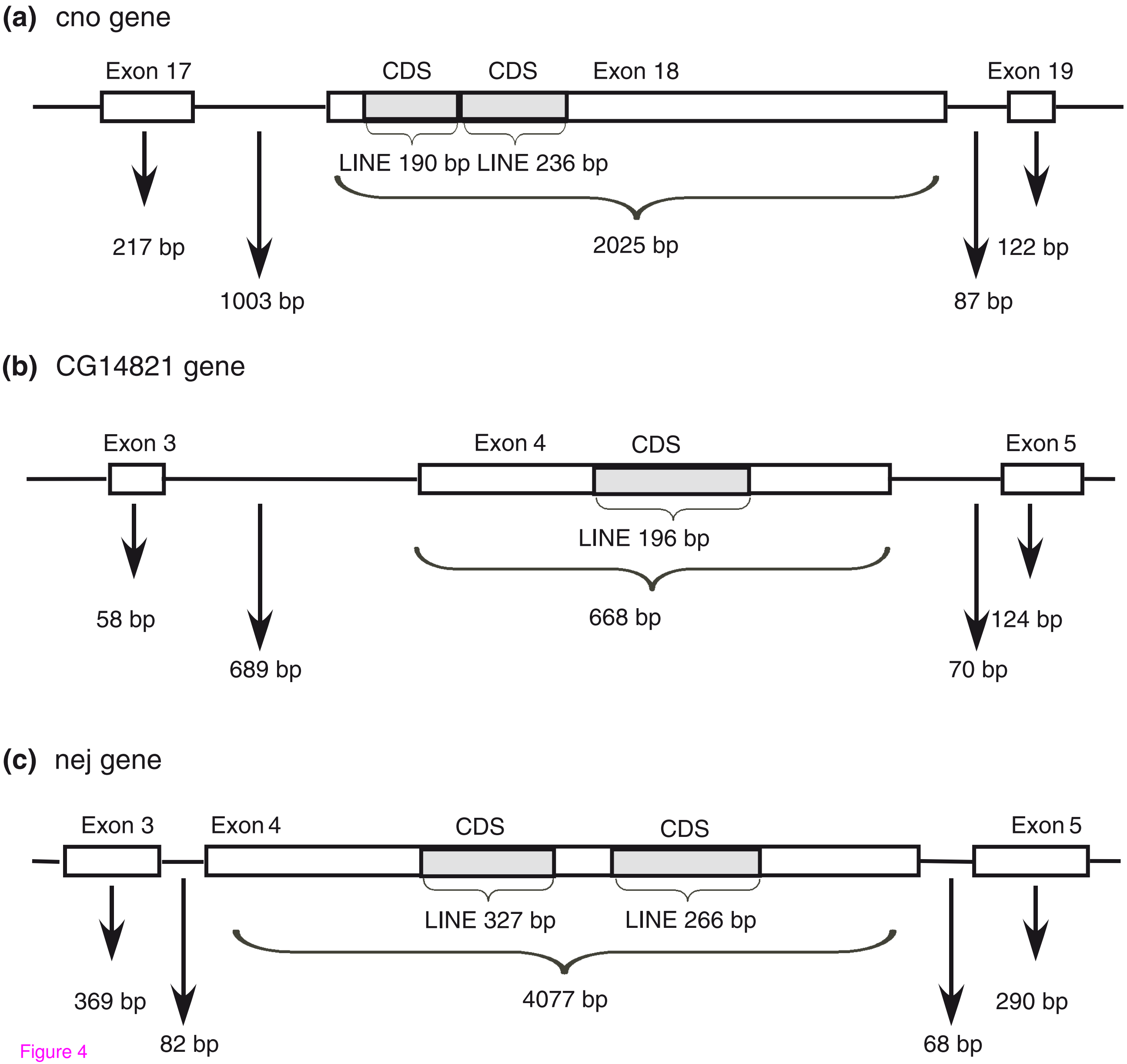

Figure 4

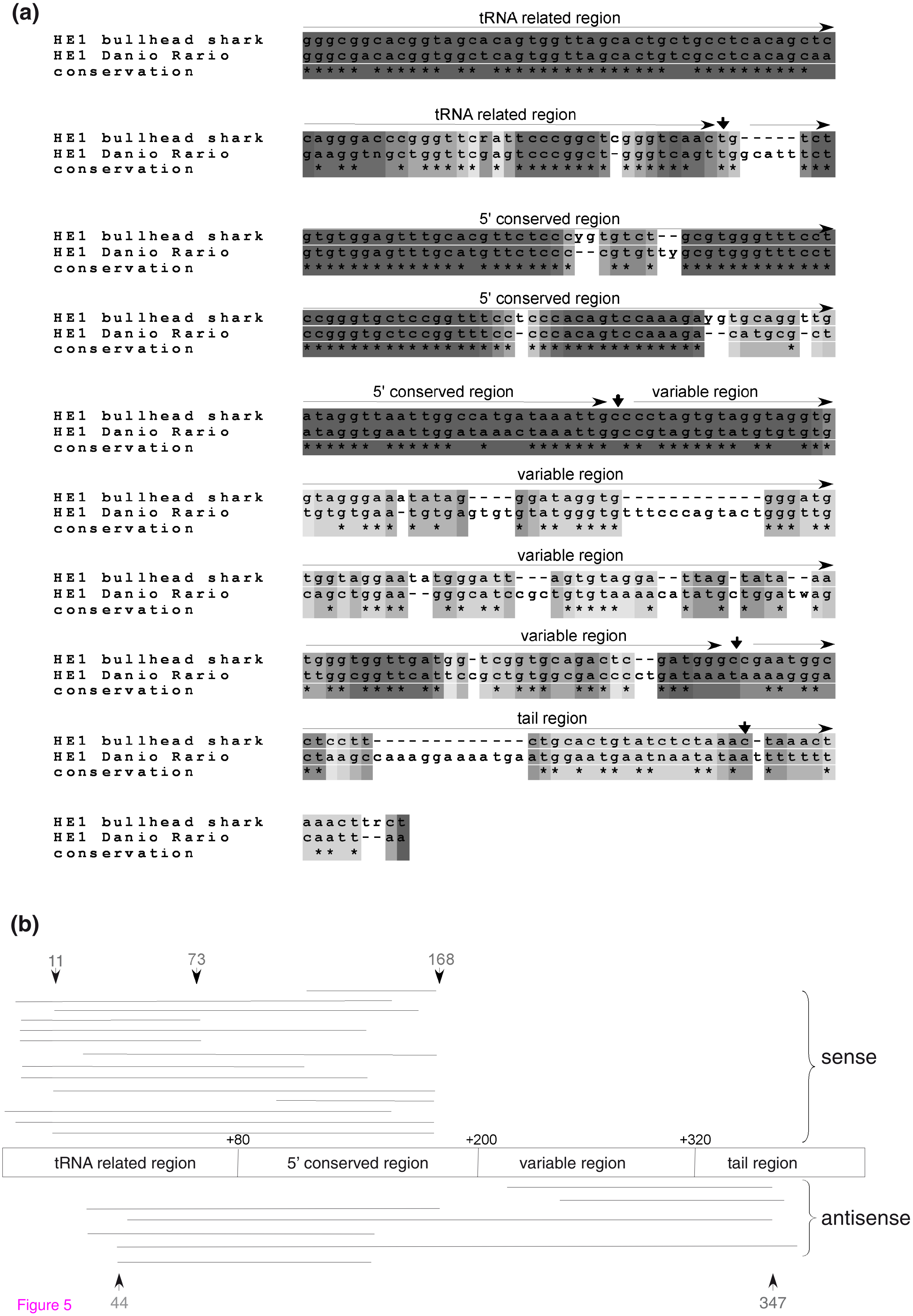
(a)
tRNA related region
HE1 bullhead shark gggcggcacggtagcacagtggttagcactgctgcctcacagctc
HE1 Danio Rario gggcgacacggtggctcagtggttagcactgtcgcctcacagcaa
conservation
tRNA related region
HE1 bullhead shark cagggacccgggttcrattcccggctcgggtcaactg-----tct
HE1 Danio Rario gaaggtngctggttcgagtcccggct-gggtcagttggcatttct
conservation
5' conserved region
HE1 bullhead shark gtgtggagtttgcacgttctcccygtgtct--gcgtgggtttcct
HE1 Danio Rario gtgtggagtttgcatgttctccc--cgtgttygcgtgggtttcct
conservation
5' conserved region
HE1 bullhead shark ccgggtgctccggtttcctcccacagtccaaagaygtgcaggttg
HE1 Danio Rario ccgggtgctccggtttcc-cccacagtccaaaga--catgcg-ct
conservation
5' conserved region
variable region
HE1 bullhead shark ataggttaattggccatgataaattgcccctagtgtaggtaggtg
HE1 Danio Rario ataggtgaattggataaactaaattggccgtagtgtatgtgtgtg
conservation
variable region
HE1 bullhead shark gtagggaaatatag----ggataggtg------------gggatg
HE1 Danio Rario tgtgtgaa-tgtgagtgtgtatgggtgtttcccagtactgggttg
conservation
variable region
HE1 bullhead shark tggtaggaatatggatt---agtgtagga--ttag-tata--aa
HE1 Danio Rario cagctggaa--gggcatccgctgtgtaaaacatatgctggatwag
conservation
variable region
HE1 bullhead shark tgggtggttgatgg-tcggtgcagactc--gatgggcgaatggc
HE1 Danio Rario ttggcggttcattccgctgtggcgaccccctgataaataaaaggga
conservation
tail region
HE1 bullhead shark ctcctt-------------ctgcactgtatctctaaac-taaact
HE1 Danio Rario ctaagccaaaggaaaatgaatggaatgaatnaatataatttttttt
conservation
HE1 bullhead shark aaacttrct
HE1 Danio Rario caatt--aa
conservation
(b)
11
73
168
sense
+80
+200
+320
tRNA related region
5' conserved region
variable region
tail region
antisense
44
347

Figure 5

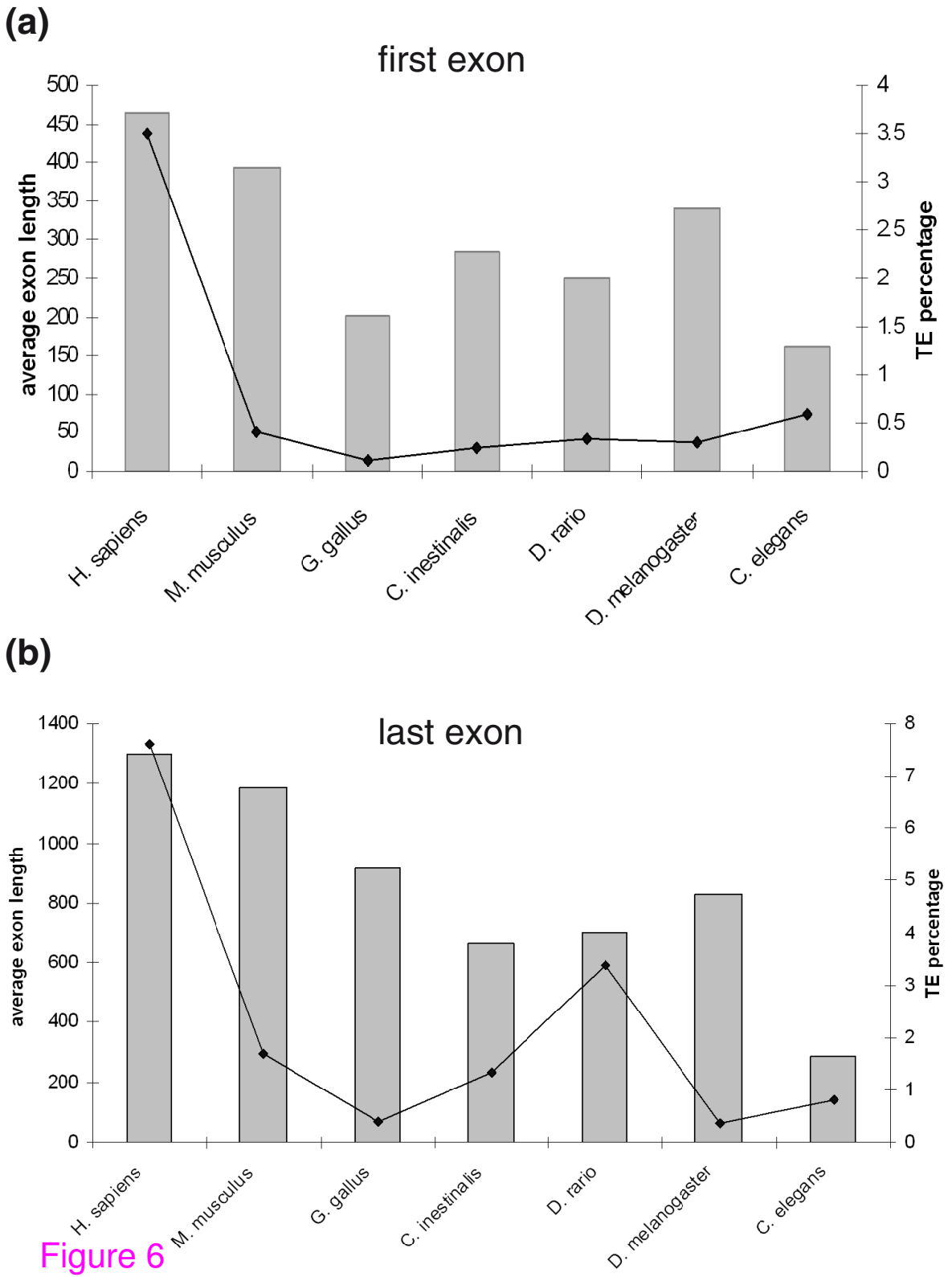

Figure 6

**Additional files provided with this submission:**

Additional file 1: Table S1.doc, 55K
http://genomebiology.com/imedia/1525618400368484/supp1.doc
Additional file 2: Table S2.doc, 27K
http://genomebiology.com/imedia/1309074752368484/supp2.doc
Additional file 3: Table S3.doc, 26K
http://genomebiology.com/imedia/1351015203368484/supp3.doc
Additional file 4: Additional data 4.doc, 27K
http://genomebiology.com/imedia/2031870622401653/supp4.doc
Additional file 5: Additional data 5.doc, 63K
http://genomebiology.com/imedia/1439336664401653/supp5.doc
Additional file 6: Additional data 6.doc, 27K
http://genomebiology.com/imedia/1128600809401653/supp6.doc